\documentclass{article}
\usepackage{amsmath}

\usepackage[super,square,sort&compress,numbers]{natbib}
\usepackage{graphicx}

\title{Statistical dynamics of social distancing in SARS-CoV-2 as a differential game}
\author{Chris von Csefalvay\thanks{Starschema Inc., Arlington, VA. Correspondence: \texttt{csefalvayk@starschema.net}.}}

\begin{document}

\maketitle

\begin{abstract}
    The novel coronavirus SARS-CoV-2 has rapidly emerged as a significant threat to global public health, in particular because -- as is not uncommon with novel pathogens -- there is no effective pharmaceutical treatment or prophylaxis to the viral syndrome it causes. In the absence of such specific treatment modalities, the mainstay of public health response rests on non-pharmaceutical interventions (NPIs), such as social distancing. This paper contributes to the understanding of social distancing against SARS-CoV-2 by quantitatively analysing the statistical dynamics of disease propagation as a differential game, and estimating the relative costs of distancing versus not distancing, identifying marginal utility of distancing based on known population epidemiological data about SARS-CoV-2 and concluding that unless the costs of distancing vastly exceed the cost of illness per unit time, social distancing remains a dominant strategy. These findings can assist in solidly anchoring public health responses based on social distancing within a quantitative framework attesting to their effectiveness.
\end{abstract}

\section{Introduction} 
\label{sec:introduction}
Where an infectious disease is not amenable to population-level prevention through vaccination and risks are non-trivial, non-pharmaceutical interventions (NPIs) remain the principal tool of public health to respond to an outbreak. This is the case with novel infectious diseases that have no specific treatment and no prophylactic (vaccine) available. In the absence of pharmaceutical interventions of proven effectiveness, in particular prophylactically, the main public health response to the emerging pandemic of COVID-19, a viral syndrome caused by the (+)ssRNA virus SARS-CoV-2 (order \emph{Nidovirales}, family \emph{Coronaviridae}, genus \emph{Betacoronavirus}, subgenus \emph{Sarbecovirus}), has rested principally on NPIs.\cite{McCoy_2020,Lai_2020,flaxman2020report,ferguson2020report} At their core, all NPIs share a quintessential relationship to social distancing, either through directly encouraging social distancing, limiting transmission potential by reducing public facilities for such encounters that may transmit the pathogen ('lockdowns'), reducing gatherings and social interactions that carry such risk ('large-gathering bans'), suspending economic activities that inherently carry the risk of social interactions or modifying the framework of activities to reduce such interactions (e.g. transitioning to remote work).

From the perspective of game theory, social distancing can be viewed as a non-cooperative game of a population $P_{1 \ldots n}$ of size $n$, where at any given time $t \in [t_0, t_f]$, each player $p_i$ adopts the strategy $\sigma(p_i, t)$. For the sake of simplicity, we will assume two fundamental axioms about social distancing behaviours:

\begin{enumerate}
	\item There exist two strategy choices, $\delta$ and $\lnot \delta$ (distancing and not distancing, respectively). At any given time, any agent $p_i \in P$ can opt exclusively for one of these two options, i.e. the strategy set $\mathcal{S}(p_i)$ is the set $\{\delta, \lnot \delta\}$.
	\item Where a player $p_i$ makes a strategy choice $\sigma(p_i, t) \in \mathcal{S}(p_i)$ at time $t$, they implement it perfectly, i.e. the efficiency of every player $p_i \in P$ in implementing their strategy $\sigma(p_i, t)$ is equal.
\end{enumerate}

Then, for the entire population $P$, the aggregate population level strategy $\bar{\sigma}(P, t)$ can be described as the sum of all strategies $\sigma(p_i, t) | p_i \in P$. We can then assign a value $\hat{\sigma}$ to each strategy, whereby the strategy of distancing, $\delta$, is assigned the value of 1 and the strategy of not distancing, $\lnot \delta$, the value of 0. 

For any given discrete time $t \in [t_0, t_f]$, where $t_f$ would be an endpoint (such as eradication, elimination, natural extinction of the pathogen, the availability of a vaccine or a combination thereof), we may then define two disjoint sets, $P_{\delta}(t)$ and $P_{\lnot \delta} (t)$, where 

\begin{equation}
	\begin{aligned}
		\forall p_j \in P_{\delta}(t), & \ \ \sigma(p_j, t) = \delta						\\
		\forall p_k \in P_{\lnot \delta}(t), & \ \ \sigma(p_k, t) = \lnot \delta		
	\end{aligned}
\end{equation}

\noindent so that $|P_{\delta}(t)| + |P_{\lnot \delta}(t)| = |P| = n$. Then, for any $t$ in discrete time $t \in [t_0, t_f]$, we define the overall factor of social distancing for a population $P$ at time $t$, $\delta(P, t)$

\begin{equation}
	\begin{aligned}
		\delta(P, t) = \frac{P_{\delta}(t)}{n} = 1 - \frac{P_{\lnot \delta}(t)}{n}
	\end{aligned}
\end{equation}

\noindent which can also be expressed as a function of strategies $\sigma(p_{1 \ldots n})$ as

\begin{equation}
	\begin{aligned}
		\delta(P, t) = \frac{\sum_{i=1}^n \hat{\sigma}(p_i, t)}{n}
	\end{aligned}
	\label{eq:delta_value}
\end{equation}

In other words, the overall factor of social distancing at time $t$, $\delta(P, t)$ is the average of the value the function $\hat{\sigma}$ assigns to each player's strategy over the entire population $P$. For simplicity's sake, we will assume that the decision process takes the shape of a continuous and simultaneous game in discrete time, and agents can instantly switch strategies with no cost (other than the cost of the strategy itself). It then holds that

\begin{enumerate}
	\item A player $p_i \in P$ opting for strategy $\sigma_{\delta}$ (social distancing) will incur $c_d$, the immediate costs of distancing. These may be social (lessened social interaction), psychological (lessened access to support systems), economic (lower access to facilities to earn) or simple matters of convenience (access to amenities). While $c_d$ is somewhat dependent on $\delta(P, t)$ (thus not distancing does not yield a benefit to a lone player in terms of access to amenities if all of these closed due to widespread social distancing), it can be assumed to be largely constant.
	\item Compared to a person opting for strategy $\sigma_{\delta}$, a person $p_i \in P$ opting for strategy $\sigma_{\lnot \delta}$ will not incur the fixed cost $c_d$, but will instead incur a relative additional cost $r_i(P, t) c_i$, where $c_i$ denotes the constant cost of illness and $r_i(P, t)$ is the risk of contracting illness when not distancing, a function of $\delta(P, t)$.
\end{enumerate}

This allows us to identify the cost function $J(\sigma, t)$ for any $\sigma \in \{\delta, \lnot \delta\}$ at time $t$ for each individual $p_i \in P$ as

\begin{equation}
	\begin{aligned}
		J(\sigma_{\delta}, t) & = c_d 									\\
		J(\sigma_{\lnot \delta}, t) & = r_i(P, t) c_i
	\end{aligned}
	\label{eq:pure_time_dependent_costs}
\end{equation}

Then, for every population level aggregate strategy $\bar{\sigma}(P, t)$ associated with the social distancing value $\delta(P, t)$ in the way described in Equation~\eqref{eq:delta_value}, the overall social cost of $\bar{\sigma}(P, t)$ can be conceptualised as

\begin{equation}
	\begin{aligned}
		\bar{J}(P, t) = n \Big( \delta(P, t) c_d + (1-\delta(P, t)) r_i(P, t) c_i \Big)
	\end{aligned}
	\label{eq:social_time_dependent_costs}
\end{equation}

For the entire time space $[t_0, t_f]$, the total social cost is then

\begin{equation}
	\bar{\bar{J}}(P) = \sum_{t_i = 0}^{t_f - t_0} J(P, t_i) = n \sum_{t_i = 0}^{t_f - t_0} \Big( \delta(P, t_i) c_d + (1 - \delta(P, t_i)) r_i(P, t_i) c_i \Big)
\end{equation}

The principal concern of this paper is not with individual action but with analysis of decision strategies on a population level. This paper will in the following conceptualise infectious disease in a population as a differential game over a differential equation form of the compartmental model first described by Kermack and McKendrick\cite{kermack1927contribution}. This model has been widely adapted and adopted since its publication in 1927,\cite{vstvepan2007kermack,roberts1999kermack,capasso1978generalization} and building on it, we will go on to identify within the statistical dynamics of that differential game the equilibria that govern ideal societal decision-making.


\section{Methods} 
\label{sec:methods}

\subsection{The ordinary differential equations of disease dynamics} 
\label{sub:the_ordinary_differential_equations_of_disease_dynamics}

Given a population of $n$ under the assumption that reinfection is sufficiently rare so as to be ignored at a population level, as appears to be the case for SARS-CoV-2,\cite{edridge2020human,deng2020primary,Bao:2020aa}), the dynamics of subpopulations with respect to infection can be modelled as a system of ordinary differential equations

\begin{equation}
	\begin{aligned}
		\frac{dS}{dt} &= - \frac{\beta S I}{n} 								\\
		\frac{dI}{dt} &= \frac{\beta S I}{n} - \gamma I 					\\
		\frac{dR}{dt} &= \gamma I
	\end{aligned}
	\label{eq:sir_equation}
\end{equation}

\begin{figure}
	\includegraphics[width=0.8\linewidth]{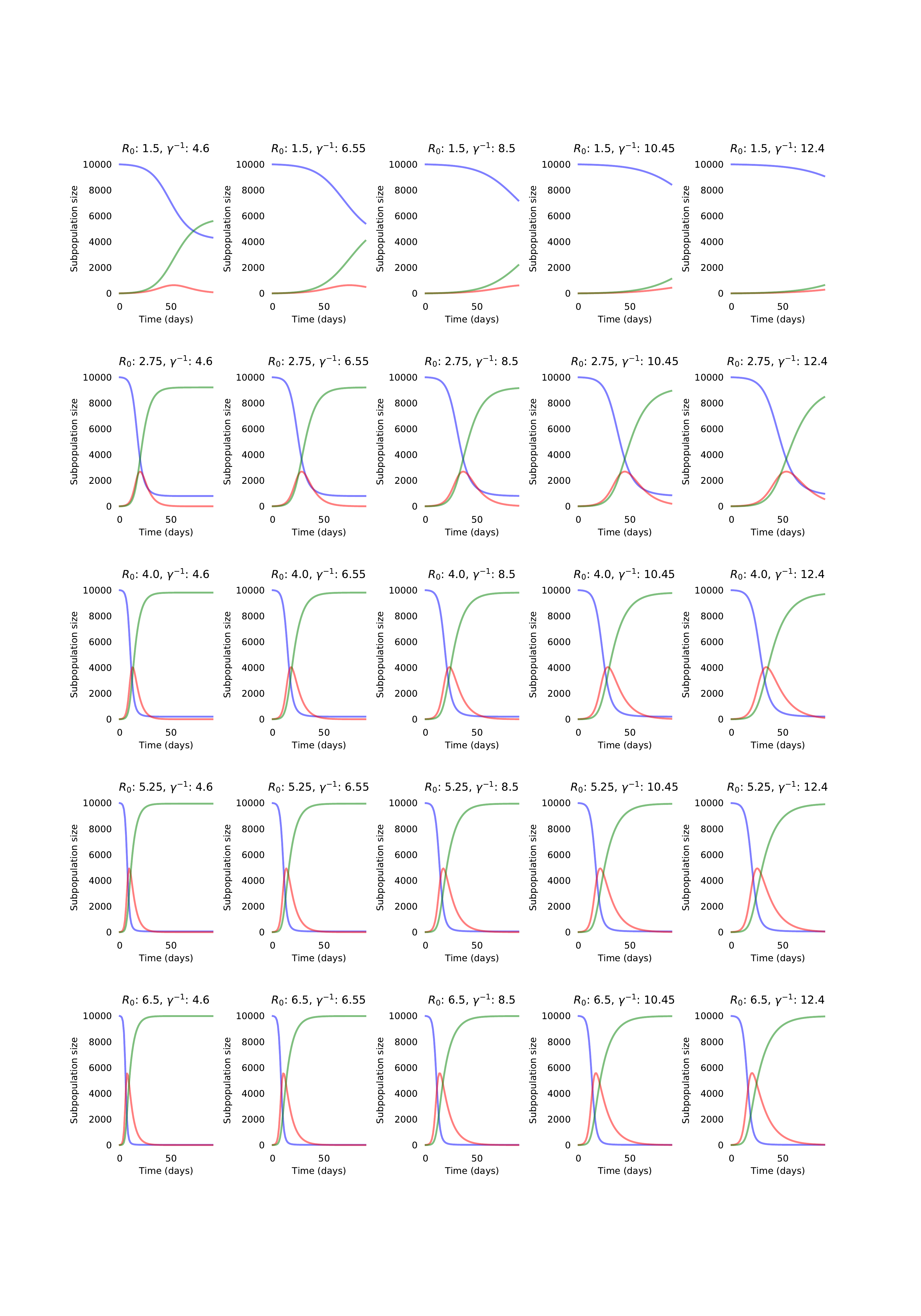}
	\centering
	\caption{Some quantitative solutions for the SIR model's population dynamics over values of $R_0$ between 1.5 and 6.5, and values of $\gamma^{-1}$ between $4.6$ and $12.4$ over a base population of 10,000 and a seed population of 0.1\% infected initially. For each plot, $\beta$ is inferred from $R_0$ and $\gamma$ using Equation~\eqref{eq:r0_equation}. The susceptible population is displayed in blue, while infected/infectious cases are marked in red and removed cases in green.}
	\label{fig:ode_solutions}
\end{figure}

\noindent under the assumption of a closed, static population, i.e. neglecting for the time being the vital dynamics (birth, unrelated death, migration) of the population. Thus, $S + I + R = n$, where $S$ represents susceptible individuals, $I$ represents infected/infectious individuals and $R$ accounts for removed individuals (mortality and recovery to immunity). In addition, due to the closed population assumption, 

\begin{equation}
	\frac{dS}{dt} + \frac{dI}{dt} + \frac{dR}{dt} = 0
\end{equation}

Furthermore, the factors $\beta$ and $\gamma$ in Equation~\eqref{eq:sir_equation} relate to each other as

\begin{equation}
	R_0 = \frac{\beta}{\gamma} | \gamma > 0
	\label{eq:r0_equation}
\end{equation}

The fraction $\frac{\beta}{\gamma}$ equals the basic reproduction number, $R_0$. For SARS-CoV-2, estimates of $R_0$ range from 1.4 to 6.49, with studies that relied on statistical estimation of $R_0$ ranging from 2.20 to 3.58, with an average of 2.67\cite{liu2020reproductive} $\gamma$, on the other hand, can be estimated as the inverse of the average number of days of illness ($\gamma^{-1}$, sometimes also described as $\tau$). This value $\tau$ has been identified by studies of the initial infection dynamics of SARS-CoV-2 to be approximately $8.5 \pm 3.9$ days.\cite{pan2020clinical,liu2020risk} Even in the absence of firm evidence as to whether SARS-CoV-2 infection followed by recovery would engender lifelong immunity or not,\cite{roy2020covid,ota2020will,lin2020duration} it can be assumed in the short term -- based on evidence from MERS-CoV and SARS-CoV -- that at least in the immediate aftermath of disease and recovery, survivors remain immune,\cite{prompetchara2020immune} and consequently $\frac{dR}{dt} \geq 0$ for any $t \in [t_0, t_f]$, i.e. the number of removed individuals ($R$) is strictly monotonously increasing over time. The inverse is true, for the same reasons, for $\frac{dS}{dt}$ and the number of susceptible individuals ($S$). Numerical solutions to this system of differential equations have been calculated using \texttt{odepack} via \texttt{SciPy 1.5.1}\cite{virtanen2020scipy} on Python 3.6, and are described in Figure~\ref{fig:ode_solutions} describes some analytical solutions for the differential equations of Equation~\eqref{eq:sir_equation} over a range of plausible values of $R_0$ and $\gamma$, with $\beta$ inferred from $\gamma$ and $R_0$ through the relationship described in Equation~\eqref{eq:r0_equation}.


\subsection{Population strategy contingent solutions to population dynamics} 
\label{sub:population_strategy_contingent_solutions_to_population_dynamics}

\begin{figure}
	\includegraphics[width=\linewidth]{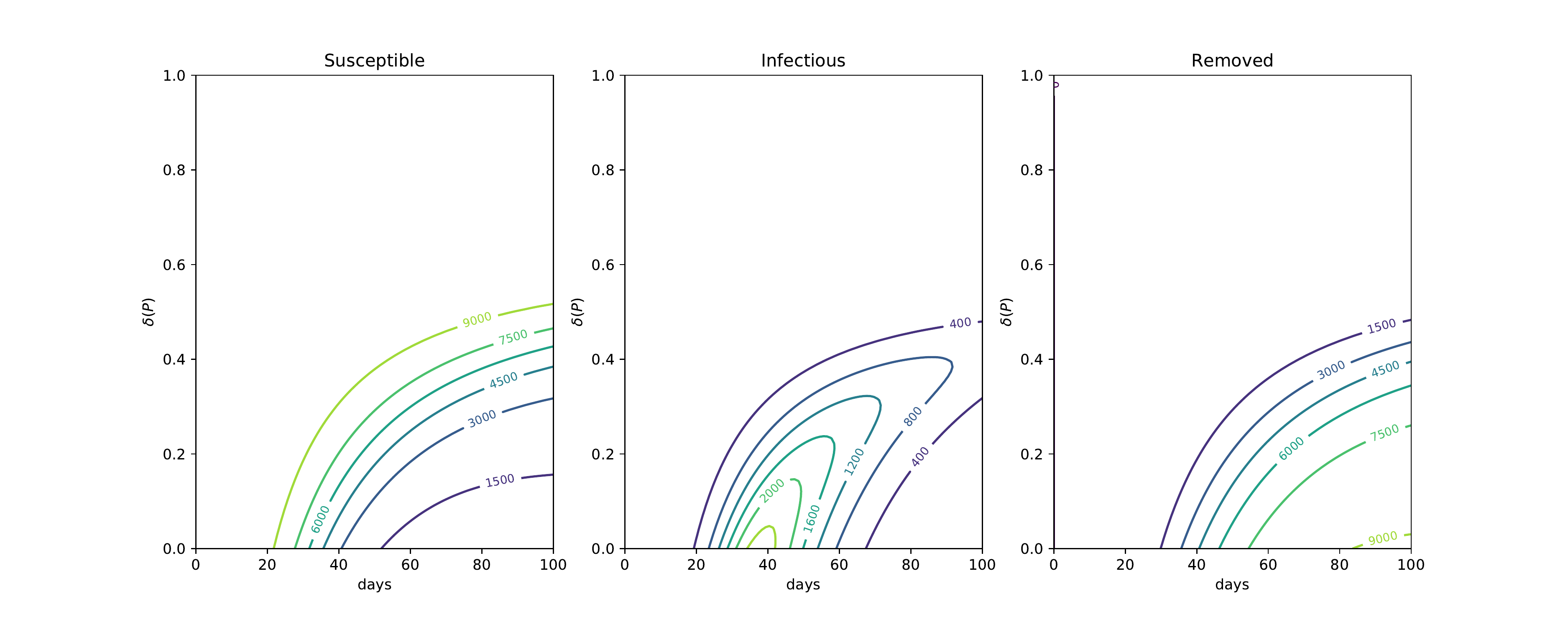}
	\caption{Contour diagrams of numeric solutions for susceptible, infectious and removed ($S$, $I$ and $R$) compartment sizes for a base population of 10,000 individuals with a seed population of 0.1\% infected, under the assumption of an $R_0$ of 2.67 and $\gamma$ of $\frac{1}{8.5}$. As this figure indicates, the 'critical mass' of social distancing takes place in the $\delta$ range of 0 to 0.4, and thus even modest increases in social distancing participation at low levels can make a significant difference in the number of infectious cases.}
	\label{fig:fig3-SIR-by-delta}
\end{figure}

Given a population that then adopts an aggregate strategy $\bar{\sigma}(P, t)$ at time $t$ that results in $\delta(P, t)$ adherence (or in Reluga's terms, investment\cite{reluga2010game}) to social distancing, the flow from $S$ to $I$ is reduced by a corresponding factor. This allows us to rewrite Equation~\eqref{eq:sir_equation} so that for an aggregate strategy $\bar{\sigma}(P, t)$ yielding $\delta$, the populations can be characterised as

\begin{equation}
	\begin{aligned}
		\frac{dS}{dt} &= - \frac{\beta S I - \delta \beta S I}{n} 			\\
		\frac{dI}{dt} &= \frac{\beta S I - \delta \beta S I}{n} - \gamma I	\\
		\frac{dR}{dt} &= \gamma I
	\end{aligned}
	\label{eq:sir_with_social_distancing}
\end{equation}

\begin{figure}
	\includegraphics[width=\linewidth]{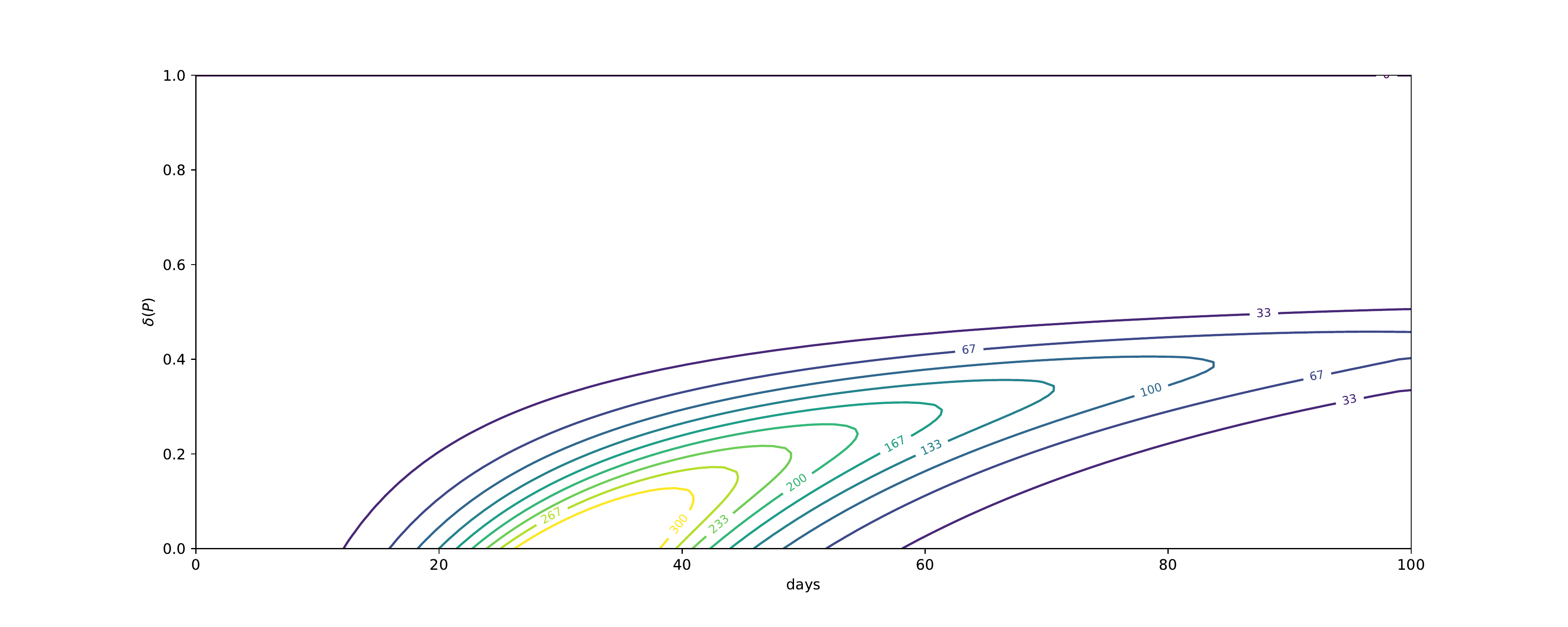}
	\caption{Marginal utility of social distancing in a population $P$ of $\delta(P)$ adherence over time. The marginal utility is defined as the vertical component of the gradient of infected individuals. The plot draws on a base population of 10,000 individuals with a seed population of 0.1\% infected, under the assumption of an $R_0$ of 2.67 and $\gamma$ of $\frac{1}{8.5}$.}
	\label{fig:marginal_utility}
\end{figure}

Solutions to this system of differential equations have been calculated and are presented in Figure~\ref{fig:fig3-SIR-by-delta}. Importantly, this allows us to identify the marginal utility $\hat{U}(P, t)$ as

\begin{equation}
	\begin{aligned}
		\hat{U}(P, t) = \frac{\partial I(P, t)}{\partial \delta(t)}
	\end{aligned}
	\label{eq:marginal_utility}
\end{equation}

\noindent i.e. the partial derivative of $I(P, t)$ over $\delta(P, t)$. Thus, for a population-level strategy $\bar{\sigma}(P, t)$ associated with $\delta(P, t)$, there exists a marginal utility function $\hat{U}(P, t)$ given $\gamma$ and $R_0$ that indicates the marginal utility at any given value of $\delta(P, t)$. This, too, can be numerically ascertained, and is shown on Figure~\ref{fig:marginal_utility}.

\subsection{Cost, risk and strategy} 
\label{sub:cost_risk_and_strategy}

Any strategy $\sigma$ has a cost $J(\sigma, t)$, as stated in Section~\ref{sec:introduction}, and the aggregate cost of $n$ individuals $p_{1 \ldots n} \in P$ each adopting, respectively, strategy $\sigma_{1 \ldots n}$, is $\sum_{i=0}^n J(\sigma_i, t)$. But since strategies are limited (one may, at any given time, either engage in social distancing or not, assuming for simplicity's sake that those who do so are entirely successful), for any aggregate strategy $\bar{\sigma} (P, t)$ resulting in a level of distancing described by $\delta (P, t)$, 

\begin{equation}
	\begin{aligned}
		\bar{J}(P, t) = \sum_{i=0}^{\delta(P, t) n} J_{\delta} + \sum_{j=0}^{(1-\delta(P, t)) n} J_{\lnot \delta}
	\end{aligned}
	\label{eq:j_bar}
\end{equation}

\noindent where $J_{\delta}$ is the cost of social distancing for discrete unit time and $J_{\lnot \delta}$ is the cost of not distancing for the same unit time. The latter of these is not constant, as Equation~\eqref{eq:pure_time_dependent_costs} shows, but a function of a constant cost of infection, $c_i$, and the risk of infection ($r_i$), which in turn is contingent on $I(t)$ and $\delta(t)$. Thus, Equation~\eqref{eq:j_bar} can be reformulated (once again, in discrete time) as

\begin{equation}
	\begin{aligned}
		\bar{J}(P, t) = \delta(P, t) n c_d + (1 - \delta(P, t)) n J_{\lnot \delta}
	\end{aligned}
\end{equation}

\noindent which expands to

\begin{equation}
	\begin{aligned}
		\bar{J}(P, t) = \delta(P, t) n c_d + (1 - \delta(P, t)) n r_i(t) c_i
	\end{aligned}
\end{equation}

For a susceptible individual $p_i \in S$, the risk of infection $r_i(t)$ in discrete time is the proportional likelihood of infection, or in other words, 

\begin{equation}
	\begin{aligned}
		r_i(t) = \frac{\beta S(t) I(t)}{n^2}
	\end{aligned}
\end{equation}

While quantification of costs of illness is difficult, quantification of the economic, social and emotional costs of social distancing is an even more complex task. However, we may, for values of $\delta(P, t)$, calculate cost fractions $\phi(P, t) = \frac{c_d}{c_i}(P, t)$, where

\begin{equation}
	\begin{aligned}
		\phi(P, t) = \frac{c_d}{c_i}(P, t) = \frac{(1 - \delta(t)) R_0 \gamma S(t) I(t)}{\delta(t) n^2}
	\end{aligned}
	\label{eq:cost_fraction}
\end{equation}

This cost fraction indicates the relative disequilibrium between the cost of distancing and the cost of illness when adjusted for risk -- in other words, for any given value of $\delta(P, t)$ at $t$, the cost fraction indicates an inflection point. As long as the cost of distancing is less than $\phi c_i$, social distancing at $\delta(P, t)$ or above is the optimum strategy. As numerical estimation of this cost fraction (Figure~\ref{fig:cost_fraction}) shows, social distancing is almost always the preferred strategy at equal cost (black line). The contour lines in Figure~\ref{fig:cost_fraction} indicate what ratio the cost of distancing has to be to the cost of illness at a given value of $\delta(P, t)$ to make distancing no longer an optimal strategy. So, for instance, a $\phi$ of 0.05 denotes the isorisk curve over $t$ where, at given $\delta(P, t)$, distancing is the preferable solution as long as its costs are less than, or equal to, 0.05 times that of illness or less. 



\section{Results} 
\label{sec:results}

\subsection{Strategies of social distancing} 
\label{sub:strategies_of_social_distancing}

As Figure~\ref{fig:fig3-SIR-by-delta} indicates for empirically ascertained values of $R_0$ and $\gamma$ based on the literature on SARS-CoV-2,\cite{Li:2020aa,liu2020reproductive} social distancing can have an overwhelmingly significant effect on the number of infectious individuals in a closed population, and the magnitude of this effect is dependent on the number of persons in the population already engaged in social distancing. This effect is most pronounced early in the epidemic (approx. 2-3 $\gamma^{-1}$ days), and the effect is most significant where less than half of the population is engaged in social distancing. Thus, unlike collective immunity in the case of vaccination, which often necessitates a fairly high level of penetration (typically estimated as $1 - R_0^{-1}$), social distancing can play a meaningful role in particular where much of the population is not yet engaged in such behaviour. This result can meaningfully inform a policy of encouraging individual social distancing early in an outbreak and dispel the misperception that marginal action is unnecessary unless a critical volume of individuals are already participating.


\subsection{Marginal utility of social distancing} 
\label{sub:marginal_utility_of_social_distancing}

Based on the key epidemiological dynamics data on the SARS-CoV-2 pandemic,\cite{Li:2020aa,liu2020reproductive} the highest marginal effect of social distancing takes place in the same early timeframe of approx. 2-3 $\gamma^{-1}$ days. Unsurprisingly, even without integrating the time-dependent discount factor proposed by Reluga (2010),\cite{reluga2010game} the numerical solutions indicate that the effect of social distancing is most significant where it is not yet a widely adopted strategy: for SARS-CoV-2, based on an initial population of 10,000 with a seed population of 0.1\% infected, the greatest marginal utility is encountered where less than 20\% of the population is engaged in social distancing, and the effect is significantly less noticeable once $\delta(P, t)$ reaches 0.5 (Figure~\ref{fig:marginal_utility}).

Calculations of marginal utility matter because they can guide public policy in determining what fraction of the population may feasibly be exempted from social distancing while still maintaining much of the benefit. Given the need for critical services, swuch as urgent medical care, emergency services and critical supply chain activities to continue even in the throes of a pandemic, marginal utility calculations based on empirical data on an outbreak may, along with the societal response (as expressed by $\delta(P, t)$, which can be empirically ascertained as well), contribute to more accurate public health measures while limiting the effect of such measures on the economy and on day-to-day life. 


\subsection{Costs and strategy choices} 
\label{sub:costs_and_strategy_choices}

\begin{figure}
	\includegraphics[width=\linewidth]{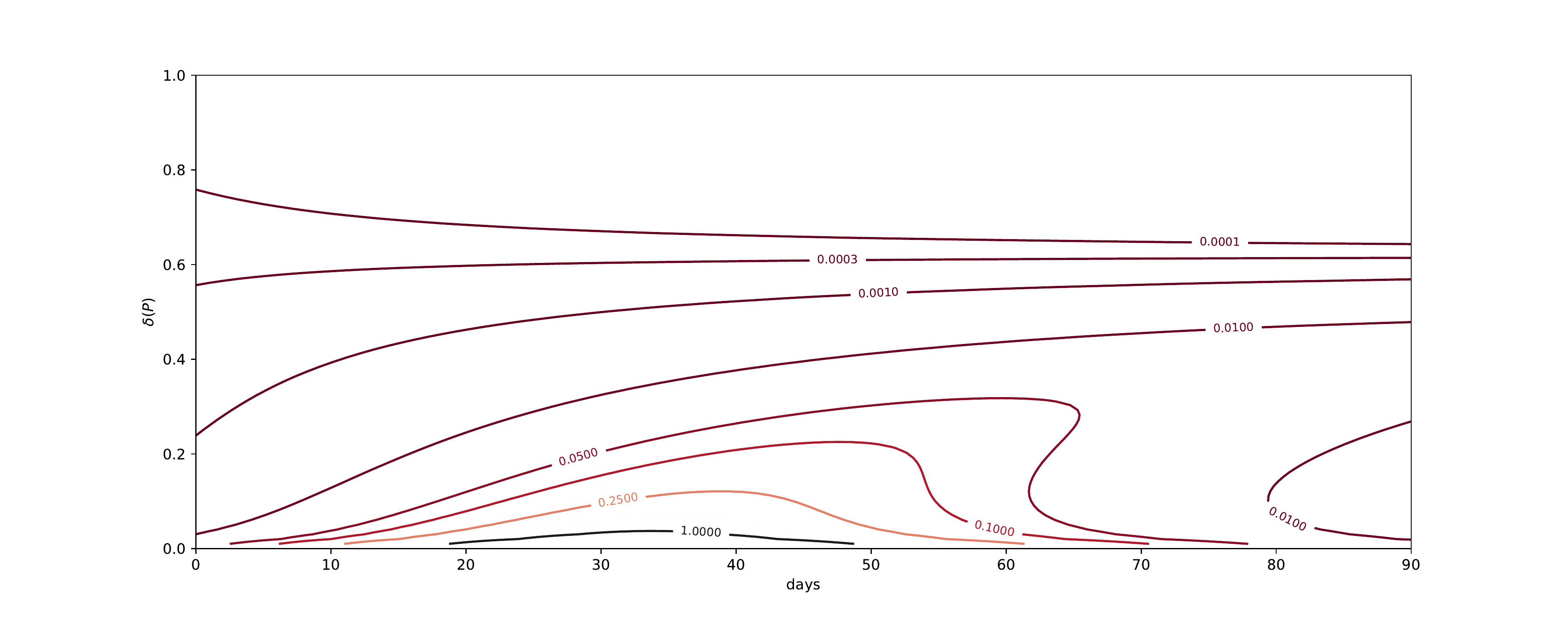}
	\caption{Cost fraction $\phi = \frac{c_d}{c_i}$ of social distancing in a population $P$ of $\delta(P, t)$ adherence over time, based on a population of 10,000 individuals with a seed population of 0.1\% infected, under the assumption of an $R_0$ of 2.67 and $\gamma$ of $\frac{1}{8.5}$. The contour lines indicate the cost fraction, i.e. what fraction of the cost of social distancing $c_d$ the cost of illness $c_i$ must be in order to make not distancing a preferred strategy.}
	\label{fig:cost_fraction}
\end{figure}

The accurate direct and indirect costs both of social distancing and of failing to do so are notoriously difficult to quantify accurately. Indeed, many of these costs are by their very nature not amenable to quantification. At the same time, by quantifying the relative ratio of cost of distancing ($c_d$) and cost of illness ($c_i$), we can identify a strategy-associated cost ratio $\phi$ that, given approximations or empirical estimates of those costs, can assist in societal decision-making with regard to social distancing. As Figure~\ref{fig:cost_fraction} shows, for most cases, the cost of social distancing would have to exceed the cost of illness by at least an order of magnitude to make it a preferable strategy. In addition, the computational solution of Equation~\eqref{eq:cost_fraction} shows not only that failure to socially distance may only be a preferable choice if the costs of distancing vastly exceed the costs of illness, but that this remains the case for much of the short term (<90 days). 

Estimates of direct medical costs of COVID-19 are difficult, but at least one study puts the median cost of a symptomatic infection at US\$3,045 in direct costs alone,\cite{bartsch2020potential} typically compounded by loss of earnings, long-term physical harm, reduction in life expectancy and quality of life and, in severe cases, the risk of mortality. The costs of social distancing are much less amenable to quantification, as these costs are primarily governed by indirect factors and intangibles, such as the cost of deferred medical treatment, second-order effects of the psychological burden inherent in decreased social interaction and the cost of lost revenue. While quantification, thus, of both the cost of distancing and cost of illness remains an outstanding subject of research, the cost fraction calculations can assist in reasoning about the best social strategy in view of these factors once ascertained or estimated.



\section{Discussion} 
\label{sec:discussion}

Pandemics pose a significant challenge to public health and social decision-making, and the COVID-19 pandemic is by no means an exception. Non-pharmaceutical interventions, such as social distancing, play a significant role in the arsenal of tools that public health authorities can bring to bear on an epidemic that is otherwise not amenable to treatment or prophylaxis. Thus, until a vaccine or a reliable therapeutic, ideally with prophylactic properties, is found, non-pharmaceutical interventions are poised to remain the mainstay of public health activity in the face of COVID-19. In view of this, an increased understanding of the way NPIs that rely on social distancing affect the statistical dynamics of SARS-CoV-2 in a population is essential for sound decisin-making.

This paper discussed a subject that is not devoid of controversy, both in the scientific and in the public realm. By their nature, NPIs interfere with citizens' day-to-day lives and may have complex economic, social and psychological effects. It is therefore important that strategy options are adequately explored from a quantitative perspective. It is hoped that in reinforcing the case for social distancing through an analysis of the statistical dynamics that underlie it, this paper can add to growing body of knowledge in support of social distancing as an effective and cost-efficient NPI where other tools are unavailable or inappropriate.


\section*{Competing interests} 
\label{sec:competing_interests}

The author declares no competing interests.


\section*{Supplementary data} 
\label{sec:supplementary_data}

All simulations, code and data are available on Github and under the DOI \texttt{10.5281/zenodo.3959666}.


\bibliography{bibliography}

\end{document}